\documentclass[a4paper,fleqn,twoside]{article}

\usepackage{espcrc2}
\usepackage{graphicx}
\newcommand{\etal}{{\it et al.}}

\title{Status and Commissioning of the CMS Experiment}

\author{
O.~Buchm{\"u}ller and F.-P.~Schilling (on behalf of the CMS collaboration)
\thanks{Talks given at the 11th Intl. Conference on B-Physics at 
Hadron Machines BEAUTY 2006, Oxford (UK), September 2006}
\address[CERN]{CERN, CH-1211 Geneva 23, Switzerland}
}
       
\begin{document}

\begin{abstract}

After a brief overview of the Compact Muon Solenoid (CMS) experiment,
the status of construction and installation is described in the first
part of the note. The second part of the document is devoted to a
discussion of the general commissioning strategy of the CMS
experiment, with a particular emphasis on trigger, calibration and
alignment. Aspects of b-physics, as well as examples for early physics
with CMS are also presented.  CMS will be ready for data taking in
time for the first collisions in the Large Hadron Collider (LHC) at
CERN in late 2007.
 
\end{abstract}

\maketitle


\section{OVERVIEW OF CMS}

\begin{figure}
\centering
\includegraphics[width=1.0\linewidth]{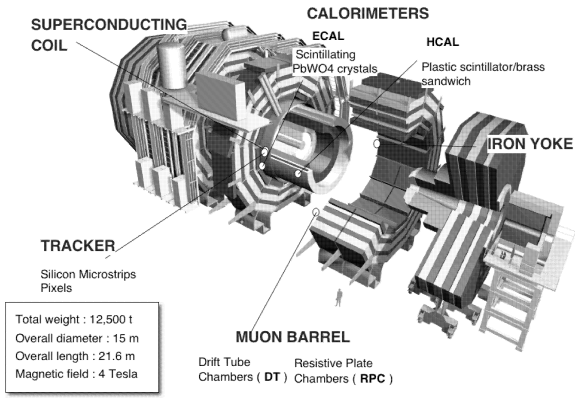}
\caption{Schematic view of the CMS detector.
\label{fig:cms}}
\end{figure}

A schematic drawing of CMS is shown in Fig.~\ref{fig:cms}. The total
weight of the apparatus is 12500 tons. The detector, which is
cylindrical in shape, has length and diameter of 21.6~m and 14.6~m,
respectively. The overall size is set by the muon tracking system
which in turn makes use of the return flux of a 13~m long, 5.9~m
diameter, 4 Tesla superconducting solenoid.  This rather high field
was chosen to facilitate the construction of a compact tracking system
on its interior while also allowing good muon tracking on the exterior.
The return field saturates 1.5~m of iron
containing four interweaved muon tracking stations.  In the central
region (pseudorapidity range $|\eta | < 1.2$) the neutron induced
background, the muon rate and the residual magnetic fields are all
relatively small, while in the forward regions ($ 1.2 < |\eta | <
2.4$) all three quantities are relatively high. As a result, drift
tube (DT) chambers and cathode strip chambers (CSC), are used for muon
tracking in the central and forward regions, respectively. Resistive
plate chambers (RPC) with fast response and good time resolution but
coarser position resolution are used in both regions for timing and
redundancy.

The bore of the magnet coil is also large enough to accommodate the
inner Tracker and the calorimetry inside. The tracking volume is given
by a cylinder of length 5.8~m and diameter 2.6~m.  In order to deal
with high track multiplicities, CMS employs 10 layers of silicon
microstrip detectors, which provide the required granularity and
precision. In addition, 3 layers of silicon pixel detectors are placed
close to the interaction region to improve the measurement of the
impact parameter of charged particle tracks, as well as the position
of secondary vertices. The electromagnetic calorimeter (ECAL) uses
lead tungstate (PbWO$_4$) crystals with coverage in pseudorapidity up
to $|\eta|<3.0$. The scintillation light is detected by silicon
avalanche photodiodes (APDs) in the barrel region and vacuum
phototriodes (VPTs) in the endcap region. A preshower system is
installed in front of the endcap ECAL for $\pi^0$ rejection. The ECAL
is surrounded by a brass/scintillator sampling hadron calorimeter (HCAL) with
coverage up to $|\eta|<3.0$. The scintillation light is converted by
wavelength-shifting (WLS) fibres embedded in the scintillator tiles
and channeled to photodetectors via clear fibres. This light is
detected by hybrid photodiodes that
can provide gain and operate in high axial magnetic fields.  This
central calorimetry is complemented by a ``tail-catcher'' in the
barrel region---ensuring that hadronic showers are sampled with nearly
11 hadronic interaction lengths.  Coverage up to a pseudorapidity of
5.0 is provided by an iron/quartz-fibre calorimeter. The Cerenkov
light emitted in the quartz fibres is detected by
photomultipliers. The forward calorimeters ensure full geometric
coverage for the measurement of the transverse energy in the event.

\section{STATUS OF CMS (SEPTEMBER 2006)}

In the current CMS Master Schedule the initial detector will be ready
for first collisions in the last quarter of 2007. Installation of the
pixel Tracker and the ECAL endcaps is foreseen during the 2007/2008
winter shutdown, in time for the first physics run in spring 2008. CMS
has made much progress over the summer. The detector has been closed
for the first time, the solenoid has been tested up to the design
field of 4~T, and cosmics data have been taken simultaneously from a
slice of all the sub-detectors using predominantly the final
components. 
The items staged for design luminosity running include the
fourth endcap muon station, RPC chambers at low angles,
several online farm slices and the third
layer of forward pixel disks.

\subsection{The Magnet}

The required performance of the muon system, and hence the bending
power, is defined by the invariant mass resolution for
narrow states decaying into muons and by the unambiguous determination
of the charge for muons with a momentum of $\approx 1$~TeV. This
requires a momentum resolution of $\Delta p/p
\approx 10\%$ at $p=1$~TeV.

To achieve this goal, CMS chose a large superconducting solenoid with
a 4~T field~\cite{magnet}. The 12.9~m long and 5.9~m diameter magnet is
operated with a current of 19.5~kA. The overall energy stored in the
field corresponds to 2.7~GJ.  The CMS magnet has been assembled on the
surface and was successfully tested during the recent CMS Magnet Test
and Cosmic Challenge (MTCC, see section~\ref{MTCC}).

\subsection{The Muon System}

Each Endcap (ME) of the Muon Detector~\cite{muon} 
contains 234 CSCs. All of the 468 endcap
CSCs have been installed on the magnet yoke disks and are now being
commissioned with cosmic rays. Each trapezoidal endcap CSC chamber has
6 gas gaps containing a plane of radial cathode strips and a plane of
anode wires parallel to the longest edge of the trapezoid and so,
roughly perpendicular to the strips.  The spatial resolution provided
by each chamber ranges from 100~$\rm \mu m$ in Station 1 to roughly
150~$\rm \mu m$ in Stations 2 to 4.  Wire signals are fast and are used
in the Level-1 Trigger though they have coarser position resolution.

The manufacture of Barrel DT chambers is complete and almost all of
the 266 DTs have been installed.  The chambers installed in the barrel yokes
(YB) are organized in 4 stations.  Each DT chamber
is piggy-backed by one or two RPCs.  The chambers are staggered from
station to station so that a high-$p_T$ muon near a sector boundary
crosses at least 3 stations. The chambers consist of twelve planes of
aluminum drift tubes; four $r\phi$ measuring planes are placed above,
and four below, a group of four $z$ measuring planes.  Each station
gives a muon vector in space with a precision of less than 100~$\rm \mu m$
in position and less than 1~mrad in direction. 
Several chambers of the barrel and endcap muon systems were
successfully operated during the MTCC
in July and August (see section~\ref{MTCC} for more details). 

\subsection{The Tracker}

The CMS Tracker~\cite{tracker} occupies a cylindrical volume of length
5.8~m and diameter 2.6~m.  The outer portion of the Tracker is
comprised of 10 layers of silicon microstrip detectors and the inner
portion is made up of 3 layers of silicon pixels. Silicon provides
fine granularity and precision in all regions for efficient and pure
track reconstruction even in the very dense track environment of high
energy jets.  The three layers of silicon pixel detectors at radii of
4, 7 and 11 cm provide 3D space points that are used to seed the
formation of tracks by the pattern recognition. The 3D points also
enable measurement of the impact parameters of charged-particle tracks
with a precision of the order of 20~$\rm \mu m$ in both the $r\phi$ and $rz$
views. The latter allows for precise reconstruction of displaced
vertices to yield efficient b tagging and good separation between
heavy and light quark jets.

The CMS Tracker continues to make good progress.  Hybrid production
was completed at the end of 2005, and module production was completed in
the spring of 2006. 
In October 2006 the forward (+) half of
the Tracker Outer Barrel TOB will be completed, Tracker Inner Barrel
TIB+ will be ready for integration into TOB+, and Tracker EndCap
TEC+ will be delivered to CERN from Aachen, where it was completed
in September.  In November 2006 the backward (-) half of TOB will be
completed, TIB- will be ready for integration into TOB- and TEC- will
be completed ready for integration into the Tracker support tube. 
The quality of the Tracker sub-detectors is very good.  The
number of dead or noisy channels is two per mille and the signal to
noise ratio is $> 25:1$.  The Pixel Detector continues to make good
progress. All components are now available and 15\% of the modules
(Barrel Pixels) and Plaquettes (Forward Pixels) have been successfully
produced. A pixel sector will be delivered to CERN in December 2006
for integration before installation into CMS in September 2007. The
full Pixel Detector will be ready to be installed into CMS in November
2007.

\subsection{Electromagnetic Calorimeter}

The ECAL~\cite{ecal} is a hermetic,
homogeneous calorimeter comprising $61\,200$ lead tungstate (PbWO$_4$)
crystals mounted in the central barrel part, closed by 7324 crystals
in each of the 2 endcaps.

About $56\,500$ of the barrel crystals have been delivered and are
being used to construct modules (400 or 500 crystals) at CERN and
in Rome. At present 122 modules out of 144 have been assembled. Thirty
bare supermodules (SM, each comprising 1700 crystals) have been
assembled. The crystal production is now proceeding at the rate of
about 1250 crystals per month. The last Barrel crystal will be
delivered by March 2007, allowing insertion of the last supermodule
in May 2007. The production of Endcap
crystals will start immediately after the end of the Barrel crystals
production and is expected to finish by February 2008. For the Barrel
electromagnetic calorimeter, the integration status is the following:
During last spring, an integration rate of 4 SMs per month was
achieved. Currently 24 SMs are completed, which is consistent with
the general construction schedule of CMS. The performance of the
integrated supermodules satisfies fully the design performance
stated in the Technical Design Report. After integration, each SM
is subjected to a pre-calibration and commissioning run of 10 days
with cosmic muons.  In addition, eight SMs have already been
pre-calibrated in an electron beam. Preliminary comparisons between
the cosmic and test beam data indicate that the cosmic data 
yield an initial inter-calibration with a precision better than
2\%. Finally, two SMs have been successfully tested in the magnet of
CMS during the MTCC, showing that the performance is maintained inside
CMS and in the 4~T magnetic field. 
Large pre-series of all off-detector
readout modules are in hand. Their production and testing will be
completed by mid-November 2006.  

\subsection{Hadronic Calorimeter}

All HCAL~\cite{hcal} module types [HB (barrel), HE (endcap), HO
(outer) and HF (forward)], including absorber and optics, are
completed.  Photodetectors and electronics have been installed and a
comprehensive calibration of HCAL using Co-60 sources has been
completed.  HF will be the first sub-detector to be lowered into the
underground experimental cavern. This is expected to occur in October
2006. The HB and HE are fully installed. The HF has been calibrated to
$\sim 5\%$, HE and HB to $\sim 4\%$. The HE and HB have also been run
globally with muons. HCAL Trigger and DAQ have been tested with
cosmics. HCAL slow controls are fully operational and data quality
monitoring (DQM) is now under development and partially up and
running.

\subsection{Trigger and Data Acquisition}

The trigger~\cite{trigger,daqtdr} system is well into production with
many components already completed. Components are exercised with other
trigger and detector electronics systems in the Electronics
Integration Center at CERN. Integration tests are run in which
detector primitives are generated and used to feed the trigger and
DAQ. Final system components were also used in data taking during the
MTCC this summer (see section ~\ref{MTCC}).

\section{THE MAGNET TEST AND COSMIC CHALLENGE (MTCC)}
\label{MTCC}

\begin{figure}
\centering
\includegraphics[width=0.99\linewidth]{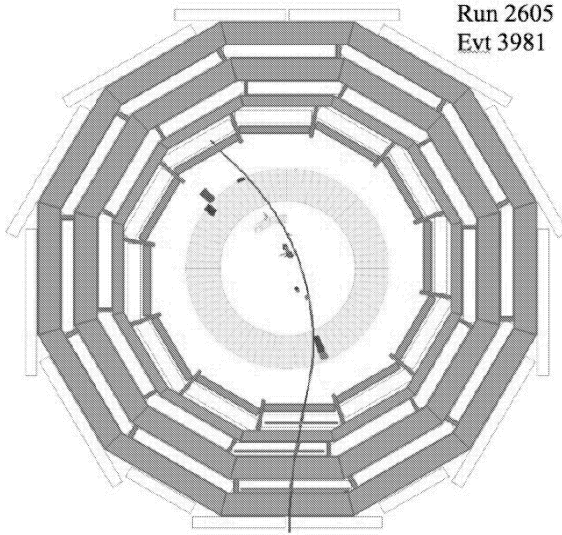}
\caption{ Characteristic event display of a cosmic muon that left signatures in all four active detector elements (Tracker, ECAL, HCAL and Muon Detector).
\label{fig:MTCC}}
\end{figure}

A full-scale test and field-mapping of the CMS 4~T solenoid magnet
system prior to lowering the major elements has been carried out this
summer. In addition to this magnet test also a ``cosmic challenge" was
performed.  The objective of this combined Magnet Test and Cosmic
Challenge was the recording, offline reconstruction and
display of cosmic muons in the 4 subsystems of CMS (Tracker, ECAL,
HCAL and Muon Detector), with the magnet operating at 4~T.

The original scope was expanded to include substantial offline as well
as online systems objectives. Data transfer to some Tier-1 centers,
online event display, quasi-online analysis at the main CERN site, and
fast offline data-checking at Fermilab were some highlighted targets
of MTCC Phase I designed to offer a first hand taste of a ``CMS-like''
running experience.

After the cabling of the detector elements the final closing of the
yoke proceeded. It was completed on 25 July allowing the start of the
magnet test. The power circuit was completed and the testing of the
coil progressed in steps: 5, 7.5, 10, 12.5, 15, 17.5 and then 19.12 kA
to reach the nominal 4~T field on 22 August. At each value, fast
discharges were provoked to learn how to tame the dumping of energy
inside the cold mass during such a discharge. After running at 3.8~T
for 48 hours for the cosmic challenge, the coil was run at 4~T for 2
hours on the 28 August, and the functional tests of the magnet have
been declared completed with success.

Twenty five million cosmic triggered events were recorded with the
principal subdetectors active, of which 15 million events have stable
field of $ \ge$ 3.8~T. Data-taking efficiency reached over 90\% for
extended periods. Several thousand of these events correspond to the
``4-detector" benchmark, and the whole data sample will provide useful
understanding and calibration of the combined detector and software
performance. A characteristic event display of a cosmic muon that left
signatures in all four active detector elements (Tracker, ECAL, HCAL
and Muon Detector) is shown in Fig.~\ref{fig:MTCC}. The shown
trajectory corresponds to the result of the standalone muon
reconstruction.  In the inner detector this trajectory coincides
within the expected uncertainties with the ECAL, HCAL and
Tracker measurements demonstrating that the propagation of the
standalone muon information was successfully carried out.  The main
conclusions of the successful MTCC are: \\
-- CMS can be opened and closed on the timescales intended; \\
-- The magnet worked stably and safely at 4~T; \\
-- The subdetectors work with the magnet and each other; \\
-- The subdetectors can be integrated with the central DAQ, trigger, DCS, DQM etc.; \\
-- The commissioning strategy broadly worked; \\
-- CMS can work as a world-wide, unified team.



\begin{table*}
\caption{Anticipated rates of $W^\pm\to\mu^\pm\nu$ and $Z^0\to\mu^+\mu^-$ events after HLT in 2008.}
\label{tab:zwrates}
\centering
\begin{tabular}{lccccc}
\hline
Luminosity & \multicolumn{2}{c}{$10^{32} \rm\ cm^{-2} s^{-1}$} &  \multicolumn{3}{c}{$2*10^{33} \rm\ cm^{-2} s^
{-1}$} \\
\hline
Time          &     few weeks  & 6 months  &   1 day   &  few weeks & one year \\
Int. Luminosity   &   $100 \rm\ pb^{-1}$ & $1 \rm\ fb^{-1}$ &   &  $1 \rm\ fb^{-1}$ &   $10 \rm\ fb^{-1}$ \\
\hline
\hline
$W^\pm\to\mu^\pm\nu$ & 700K & 7M & 100K & 7M & 70M \\
\hline
$Z^0\to\mu^+\mu^-$   & 100K & 1M &  20K & 1M & 10M \\
\hline
\end{tabular}
\end{table*}

\section{CMS COMMISSIONING STRATEGY}

Once CMS is fully installed in the underground cavern, a comprehensive
program of commissioning has to be carried out in order to optimize
the detector performance, and to prepare it for an optimal
exploitation of its physics potential.

The commissioning program includes the precise
calibration of the ECAL and HCAL
calorimeters, the alignment of the tracking and muon detectors and 
the definition of an efficient and flexible trigger setup.  Once the
detector is aligned and calibrated, physics tools such as b-tagging
and missing $E_T$ measurement can be commissioned.

\subsection{LHC Schedule}

Recently, the LHC start-up schedule has been revised~\cite{lhc}.  The
closing of the experiments will be 
followed by the start-up and initial commissioning of
the LHC accelerator with single beams at $E=450 \rm\ GeV$ in autumn 2007. 
In December
2007, a 2-3 week ``calibration run'' will be carried out with
collisions at $\sqrt{s}=900 \rm\ GeV$ and low specific luminosities
$\mathcal{L}_{sp}\sim 10^{29}
\rm\ cm^{-2} s^{-1}$.

After the winter shutdown, LHC operation will be continued with
commissioning at $7 \rm\ TeV$  and a three-stage  running scenario: 
(1) a one-month ``pilot
physics run'' at $\mathcal{L}_{sp} \sim 10^{32} \rm\ cm^{-2}s^{-1}$;
(2) pushing of the machine parameters in order to increase the
specific luminosity to $10^{33} \rm\ cm^{-2}s^{-1}$; (3) towards the
end of 2008 the luminosity could be pushed to the nominal $2*10^{33}
\rm\ cm^{-2}s^{-1}$ value.

\subsection{Commissioning Data Samples}

From the CMS commissioning point of view three distinct phases can be
identified, which give access to complementary data sets: \\
-- {\bf No beams:} Accumulation of large
samples of cosmic muons, which are very beneficial e.g. for Tracker
and muon barrel alignment. \\
-- {\bf Single beams:} Single beams give rise
to beam halo muon and beam-has events. The near-horizontal beam halo
muons can be used for Tracker and muon end-cap alignment. Beam-gas
interactions can be used for various commissioning tasks. \\
-- {\bf Colliding beams:} LHC collisions will provide
a cocktail of physics events, depending on the specific
luminosity. For commissioning purposes, muons and electrons from
$W^\pm,Z^0$ decays, but also minimum bias and QCD jet events are
useful.

During the low-luminosity 2007 calibration run at $\sqrt{s}=900 \rm\
GeV$, the available data will be completely dominated by minimum bias
and QCD jet events.  At most a few tens of $W^\pm,Z^0$ events can be
expected in this phase.  However, the 2008 pilot run will see
substantial rates of $W^\pm,Z^0$ events
(Tab.~\ref{tab:zwrates}). Large samples of high $p_T$ muons from these
events can thus be accumulated within a short timescale and used
e.g. for alignment.

\subsection{CMS Commissioning Strategy}

The main CMS commissioning goals are: (1) Efficient operation of trigger
and DAQ; (2) Tracker and Muon alignment; (3) Calorimeter calibration.
Once these goals are achieved, the commissioning
of higher level physics tools and objects such as b-tagging, jets,
missing $E_T$ can proceed.

Provided a sufficient amount of physics, cosmics and beam-halo/gas 
events can be accumulated, a lot of
important commissioning tasks can be performed already during the 2007
calibration run: Trigger and DAQ can be timed in, synchronized and their
data integrity be checked. The trigger algorithms can be debugged and
improved.  ECAL and HCAL can be calibrated to $\sim 2\%$.
Tracks from cosmic and beam halo muons as well as collision tracks can
be used to align the Tracker to significantly better than $100 \rm\
\mu m$, and to align the muon chambers.

However, important commissioning tasks can only be carried out in
2008: The final HCAL and ECAL calibrations, the final Tracker
alignment (the pixel detector will only be installed in 2008) and the
commissioning of b-tagging, missing $E_T$ etc.


\section{MAJOR COMMISSIONING TASKS}

Details on the commissioning procedures as well as the physics
potential can be found in~\cite{ptdr1,ptdr2}.  In the following, a few
aspects of the major commissioning tasks are highlighted.

\subsection{Trigger and DAQ}

The CMS trigger consists of the hardware based Level-1 trigger
(nominal accept rate $100 \rm\ kHz$) using calorimeter and muon
signals, and of the High Level Trigger (HLT), a massively parallel
processor farm in which the full event information is available. The
HLT reduces the rate by three orders of magnitude to $100 \rm\ Hz$
which are sent to the Tier-0 centre for prompt reconstruction.

An efficient operation of the trigger is ensured if both ECAL and HCAL
are calibrated to the level of $2\%$, the muon detector is aligned to
$500\rm\ \mu m$, and the silicon Tracker is aligned to $20\rm\ \mu m$
(for HLT b-tagging).  Most of these requirements can be met already
during the 2008 pilot run.

Trigger tables~\cite{daqtdr} for low and high luminosity running have
been prepared and are currently being refined. In addition, trigger
scenarios for the 2007 and 2008 calibration and pilot runs are being
prepared, with a focus on commissioning triggers.

The data reconstructed at the Tier-0 are split in $\mathcal{O}(10)$
streams~\cite{rtag}, according to their physics content. Particularly
important for commissioning purposes is the {\em express stream},
which is defined to have a fast turn-around time.  Its purpose is to
provide fast feedback on any possible interesting high mass signals,
but also to provide data sets used for alignment and calibration.


\subsection{Tracker Alignment}

The CMS Silicon Tracker consists of $\sim15000$ silicon strip and
pixel sensors covering an active area of $\sim200 \rm\ m^2$.  This
large number of independent silicon sensors and their excellent
intrinsic resolution of $10\ldots50\rm\ \mu m$ make the alignment of
the CMS Tracker a complex and challenging task.

\subsubsection{Impact of Misalignment}

\begin{figure}
\centering
\includegraphics[angle=0,width=0.835\linewidth]{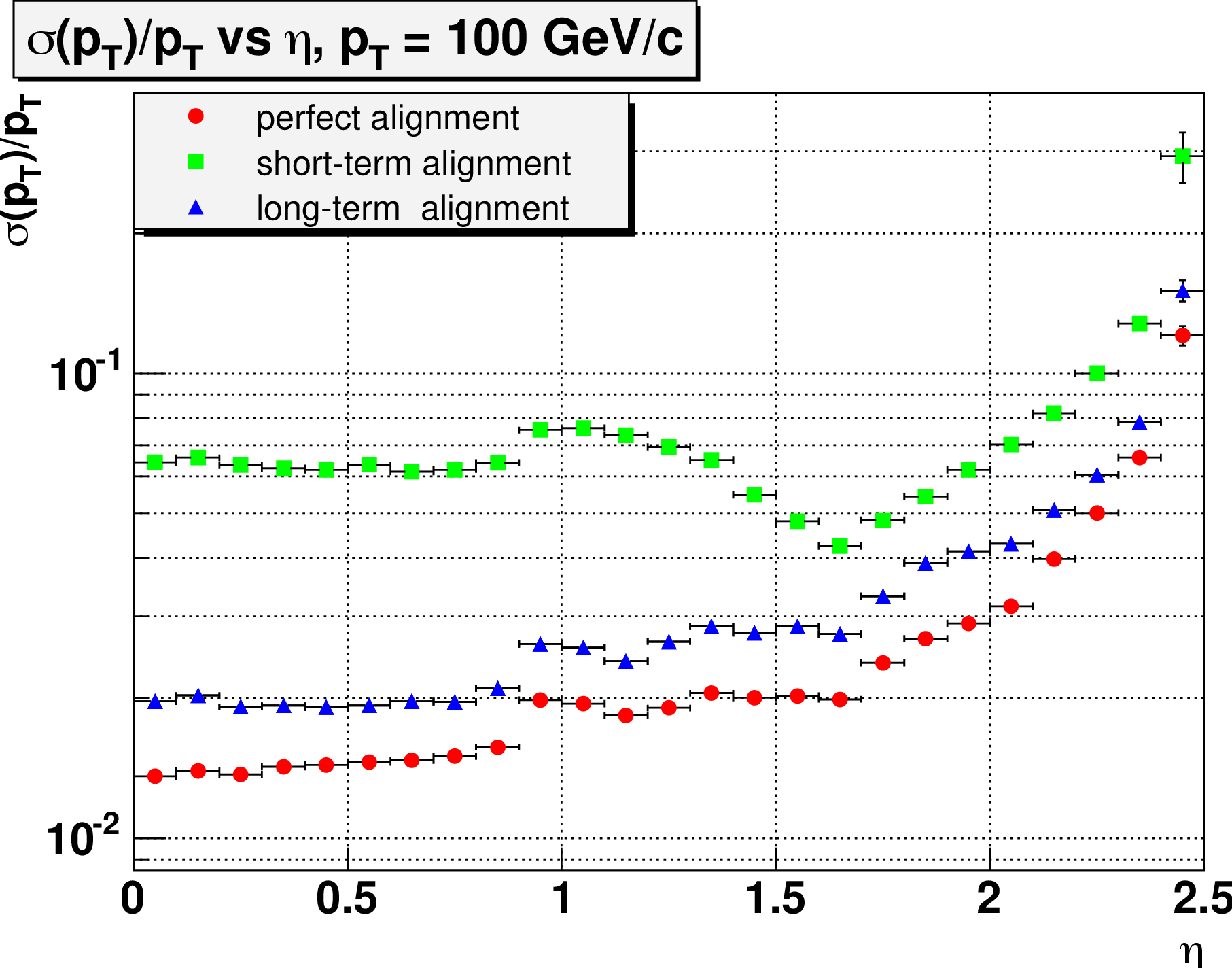}
\caption{Transverse momentum resolution in the CMS Tracker with and
without misalignment.
}
\label{fig:misali}
\end{figure}

Misalignment will degrade the track parameter resolution and hence
affect the physics performance of the Tracker, for instance the mass
resolution of resonances or the b-tagging performance.  To assess the
impact of misalignment on the tracking and vertexing performance, Two
{\em misalignment scenarios} have been
implemented~\cite{misaliscen,misalinote}, which are supposed to mimic
the conditions for different data taking conditions, namely the {\em
First Data Taking Scenario} and the {\em Long Term Scenario}
(Fig.~\ref{fig:misali}).

\subsubsection{Alignment}

The alignment strategy for the CMS Tracker foresees that in addition to
the knowledge of the module positions from measurements at
construction time, the alignment will proceed by means of a Laser
Alignment System (LAS) and track-based alignment.  The LAS uses
infrared laser beams and operates globally on the larger Tracker
composite structures. It cannot determine the positions of individual
modules.  The goal of the LAS is to provide measurements of the
Tracker substructures at the level of $100 \rm\ \mu m$ 
as well as monitoring of possible
structure movements at the level of $10 \rm\ \mu m$.

\begin{figure}
\centering
\includegraphics[angle=270,width=0.99\linewidth]{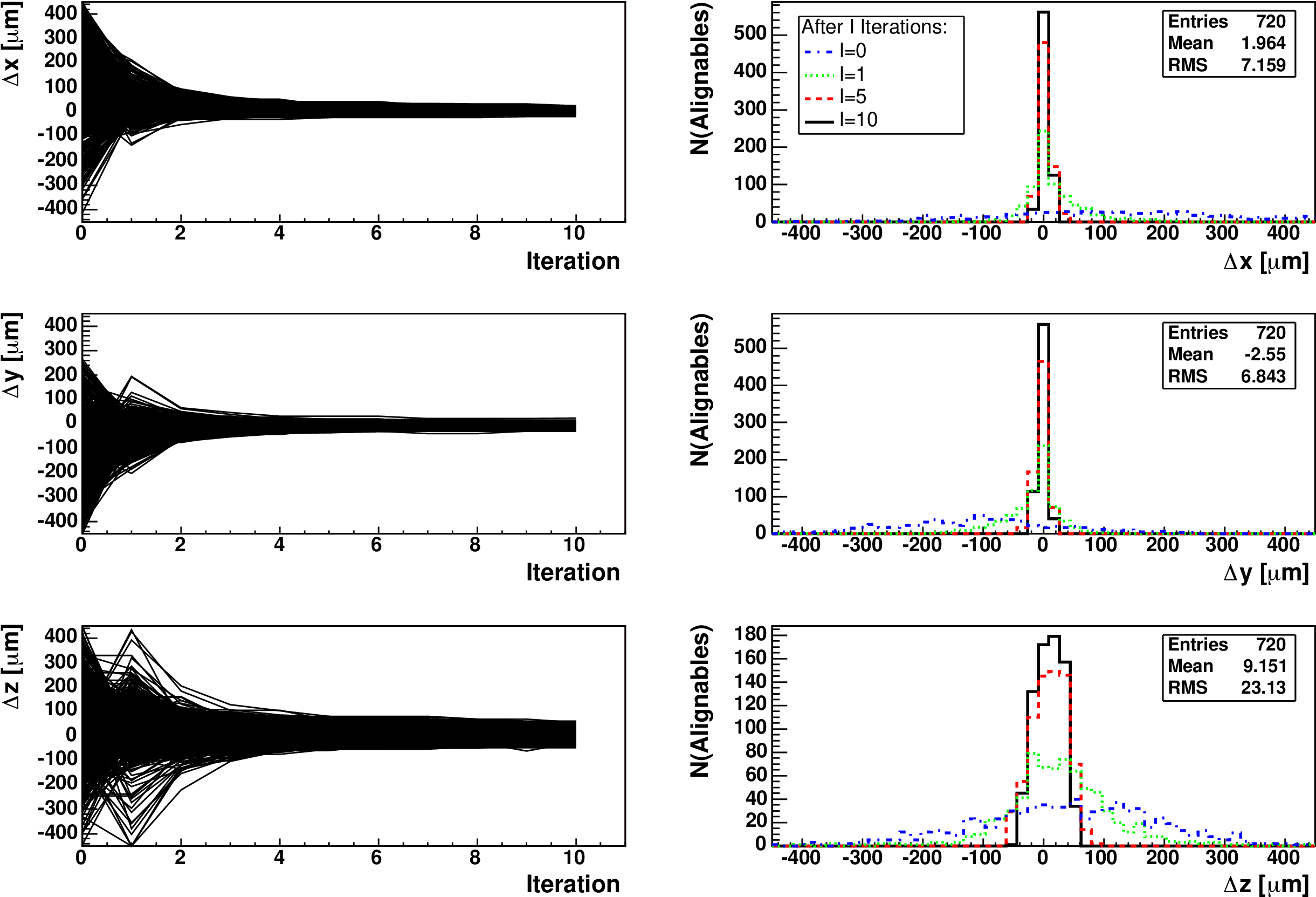}
\caption{Alignment of the CMS Pixel detector using the HIP algorithm.
}
\label{fig:hip}
\end{figure}

Track-based alignment represents a major challenge at CMS because the
number of degrees of freedom to be determined with a precision of
$\sim 10 \rm\ \mu m$ is $\mathcal{O}(100,000)$. Three different
algorithms for alignment with tracks are used in CMS: \\
-- {\bf HIP algorithm:} Minimization of 
a local $\chi^2$ function on each sensor~\cite{hipnote}.  Correlations
between different sensors are not explicitly included, but taken care
of implicitly using iterations. No inversions of large matrices are
involved (Fig.~\ref{fig:hip}). \\
-- {\bf Millepede-II:}
A global linear least-squares algorithm which takes into account
correlations among parameters.  For $N$ alignment parameters the
solution requires the inversion of a $N {\rm x} N$ matrix.  A new
version, Millepede-II, was developed~\cite{millenote} which offers
additional solution methods and is expected to be scalable to the
full CMS Tracker alignment problem within reasonable CPU time. \\
-- {\bf Kalman filter:}
A method for global alignment derived from the Kalman filter. It is
iterative and avoids inversions of large
matrices~\cite{kalmannote}. The alignment parameter update can be
extended to those elements that have significant correlations with the
ones in the current track.

The current alignment strategy foresees that cosmics and beam halo
muons can be used to carry out an initial alignment of the strip
Tracker in 2007, which could be improved using high $p_T$ collision
tracks when available.  When larger samples of muons from $W^\pm,Z^0$
decays become available in 2008, a standalone alignment of the by then
installed pixel detector can be performed, followed by the precise
alignment of the strip Tracker, using the pixel detector as a
reference system.


\subsection{Muon Alignment}

The CMS Muon system consists of 790 individual chambers with an
intrinsic resolution in the range $75\ldots100 \rm\ \mu m$.  Excellent
alignment of the Muon system is particularly important to ensure
efficient muon triggering and good track momentum resolution at large
momenta.

For optimal performance of the Muon spectrometer over the entire
momentum range up to 1 TeV, the different muon chambers must be
aligned with respect to each other and to the Tracker to within $100
\rm\ \mu m$. To control misalignment during commissioning and to
monitor further displacements during operation, CMS will combine
measurements from an optical-mechanical system with the results of
track based alignment~\cite{muonalinote}.


\subsection{Calorimeter Calibration}

\begin{figure}
\centering
\includegraphics[angle=0,width=0.9\linewidth]{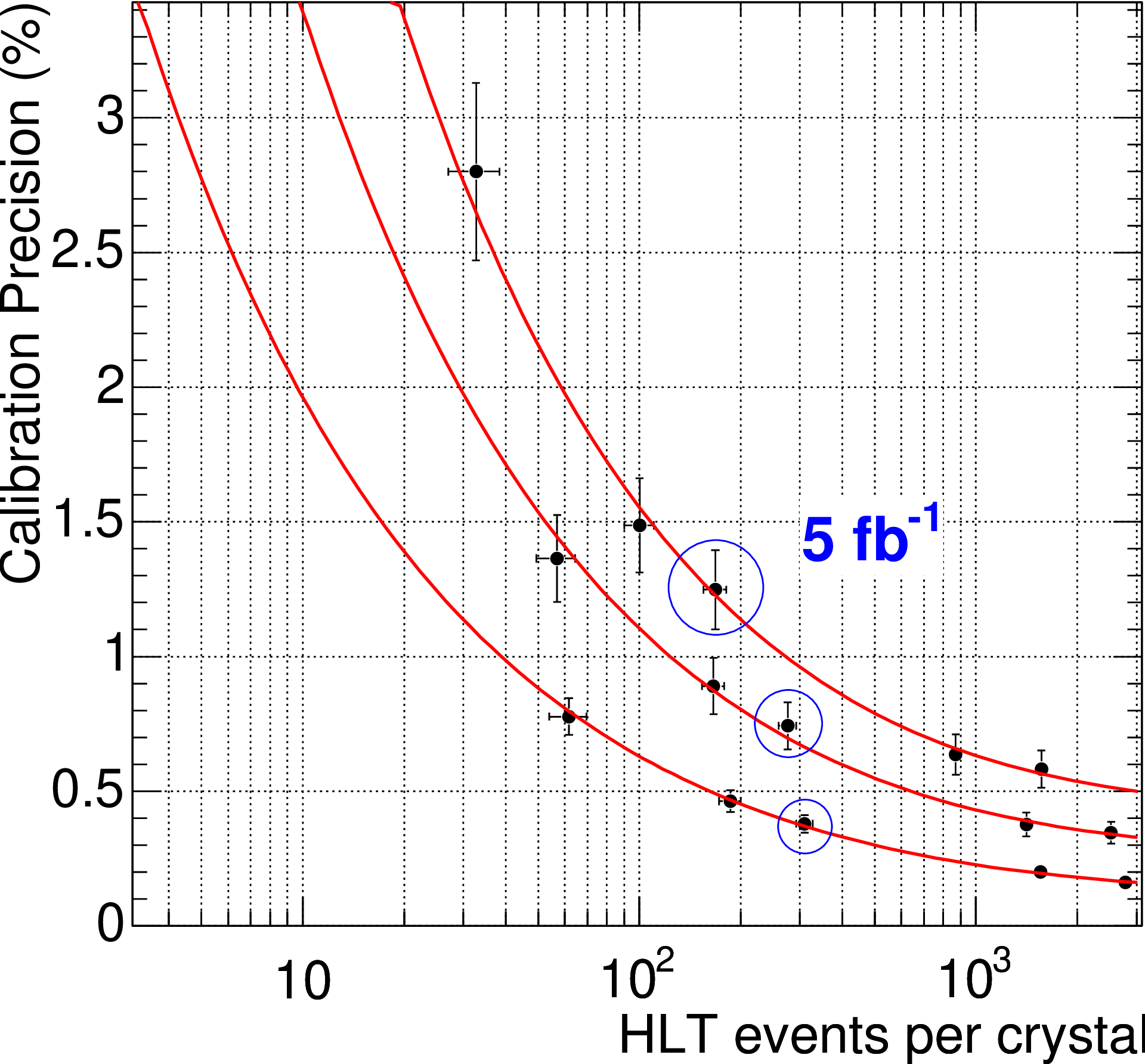}
\caption{ECAL calibration precision.}
\label{fig:ecal2}
\end{figure}

Precise calibration of the ECAL and HCAL calorimeters is a key
ingredient for precise measurements of photons, electrons, hadrons,
jets and missing $E_T$. Certain physics channels, such as
$H^0\to\gamma\gamma$, impose very tight requirements such as the ECAL
calibration being known to the level of $0.5\%$. On the other hand
typical SUSY signatures involve final states with jets and missing
$E_T$, measured using the HCAL. The knowledge of the energy scale for
b-jets is crucial for top quark mass measurements.

HCAL will be pre-calibrated to the level of $4\%$ using a radioactive
source. This calibration will be improved upon using minimum bias
events for HCAL uniformity, $E/p$ from high $p_T$ isolated tracks
extrapolated from the Tracker, and di-jet balance for regions not
covered by the Tracker acceptance~\cite{ptdr1}.

For ECAL, a pre-calibration using test beam and light-yield
measurements as well as cosmics will be performed to a precision of
around $2\%$.  During the 2007 calibration run, it is envisaged to set up a
dedicated minimum bias event stream with a bandwidth of $1 \rm\
kHz$. Using phi symmetry as well as $\pi^0/\eta^0\to\gamma\gamma$
decays, the calibration can be improved quickly.  The
ultimate ECAL calibration precision will be reached from the 2008
pilot physics run onwards, using the $E/p$ method with high momentum
electrons from $W,Z$ decays (see Fig.~\ref{fig:ecal2}).  Studies using
full simulation have shown that a precision of $\sim 0.5\%$ can be
achieved in the barrel using $\sim 5 \rm\ fb^{-1}$ of
data~\cite{ecalcalib}.


\section{EARLY PHYSICS WITH CMS}

Provided the commissioning tasks can be be performed successfully,
there is an exciting potential for early physics with
CMS~\cite{ptdr2}. Here, only two examples can be highlighted: 

\subsection{Top Physics}

Since the top quark pair production cross section is very large at the
LHC ($\sim 830 \rm\ pb$), top physics is an early physics topic for CMS.
Cross section and mass measurements will be possible in all major decay
channels. For $\mathcal{L}=1 \rm\ fb^{-1}$, $\sim 700$ events are
expected in the dilepton channel. A cross section measurement at the
$10\%$ level, as well as a mass measurement to $\Delta m_t \sim 4.2
\rm\ GeV$ will be possible~\cite{top1}, the systematic error being
dominated by the b-jet energy scale uncertainty.  At
$\mathcal{L}=10\rm\ fb^{-1}$, the top mass can be measured to $\Delta
m_t \sim 1.2 \rm\ GeV$ in the semileptonic channel~\cite{top2},
provided the b-jet energy scale is known to $\sim 1.5\%$.
These examples illustrate the importance of the commissioning of
physics tools such as b-tagging (alignment) and jet energy scale
(calibration).

\subsection{High Mass Dileptons}

Resonances at high mass in dilepton final states are very interesting
for early discoveries since they could potentially show up at
luminosities as low as a few $100 \rm\ pb^{-1}$. High mass dileptons
(where $M_{ll}\sim 1 \rm\ TeV$) are predicted by various new physics
scenarios.
Simulations show~\cite{ptdr2} that a resonance in the di-muon mass
spectrum could be discovered within a few weeks of data taking.
However, for a measurement of the mass of the resonance and to
separate it from the continuum background, a very good alignment of the
Tracker and muon detectors is crucial.


\section{ASPECTS OF B-PHYSICS}

\subsection{b-Tagging Performance}

\begin{figure}
\centering
\includegraphics[angle=0,width=0.9\linewidth]{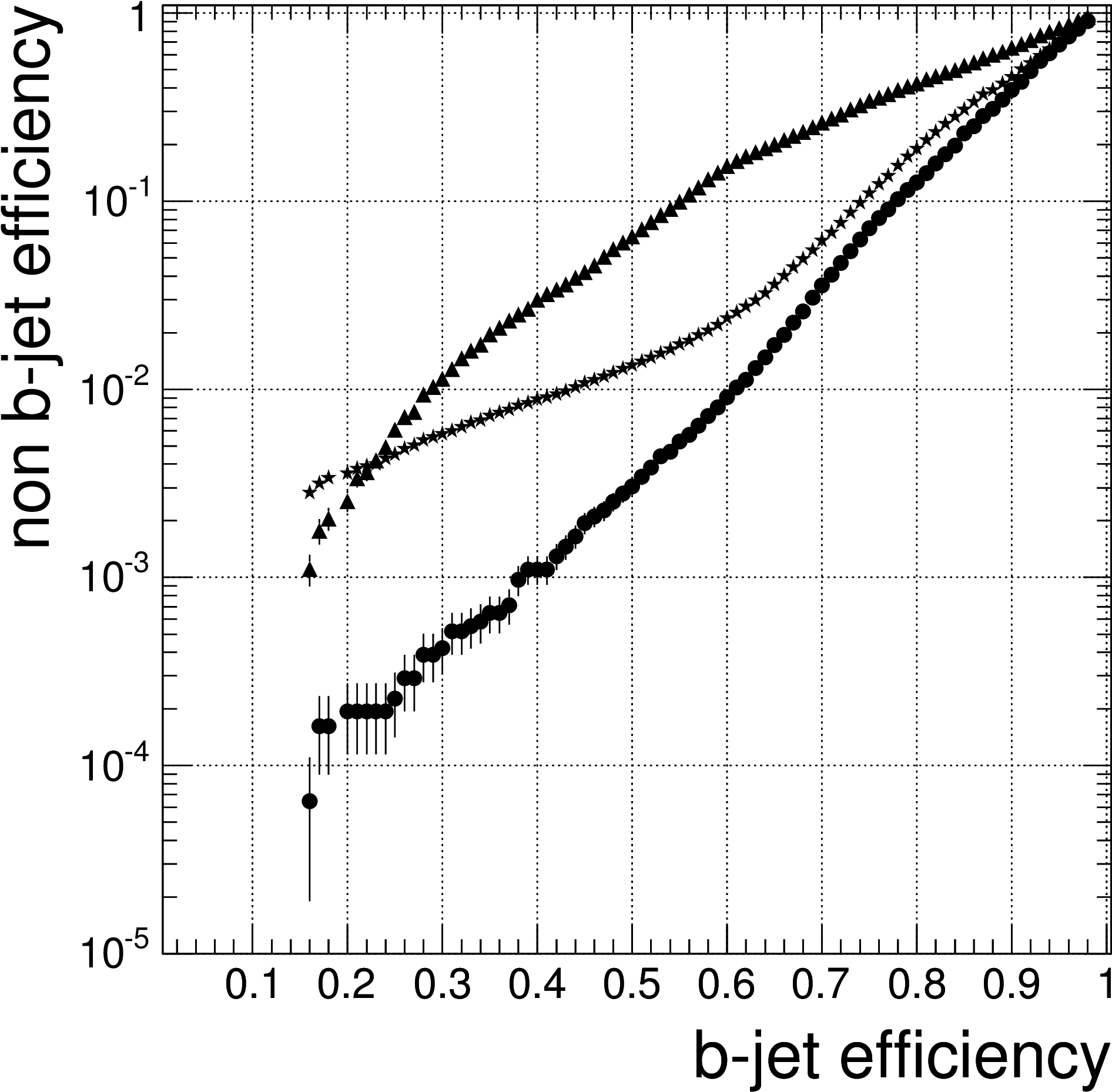}
\caption{B-tagging performance for c (top), g (middle) and uds (bottom) jets.}
\label{fig:btag}
\end{figure}

Various b-tagging algorithms have been implemented in the CMS
reconstruction software and their performance evaluated
in~\cite{ptdr1}. Both lifetime as well as soft lepton ($e^\pm$ and
$\mu^\pm$) tags have been studied.  The lifetime-tag based algorithms
are: \\
-- {\bf Track counting:} Robust algorithm which counts
the number of tracks in a jet with impact parameter above a given
threshold; \\
-- {\bf Probability:} Calculates the probability that a set of tracks  
originate from the primary vertex; \\
-- {\bf Combined secondary vertex tag:} Reconstructs
the secondary vertex of the b hadron decay and combines
several discriminating variables~\cite{btagnote}
 (Fig.~\ref{fig:btag}).

All lifetime based algorithms require that the pixel detector has been
aligned with tracks.


\subsection{Sensitivity to {\boldmath $B^0_s\to\mu^+\mu^-$}}

A new study of the CMS sensitivity to the rare decay
$B^0_s\to\mu^+\mu^-$ has been performed~\cite{bsmumunote}.  It uses a
dedicated HLT trigger (accept rate $1.7 \rm\ Hz$) and a cut-based
offline analysis. For a signal efficiency of $\sim 2\%$, a background
rejection of $2.6*10^{-7}$ is obtained. For $\mathcal{L}=10 \rm\
fb^{-1}$ this corresponds to $6.1$ $(13.8)$ signal (background)
events.  The corresponding $90\%$ C.L. upper limit on the branching
fraction is $\mathcal{B}(B^0_s\to\mu^+\mu^-) < 1.4*10^{-8}$, including
systematic errors.


\section{CONCLUSIONS}

The construction and installation of the CMS detector is making very
good progress, demonstrated for instance by the successful Magnet Test
and Cosmic Challenge carried out in the summer of 2006.  Following the
installation of CMS, the LHC 2007 calibration run and 2008 pilot
physics run must be used for detector commissioning. Important
examples are the trigger and DAQ system, Tracker and Muon alignment
and calorimeter calibration. The commissioning of physics tools relies
on the success of these tasks. There is an exciting program for early
physics with CMS.


\def\cmsnote#1#2{CMS Note {\bf #1/#2}}
\def\cmscr#1#2{CMS CR {\bf #1/#2}}


\begin{thebibliography}{99}




\bibitem{magnet}
CMS Coll., ``Magnet Technical Design Report'', CERN/LHCC {\bf 1997-010} (1997).

\bibitem{muon}
CMS Coll., ``Muon Technical Design Report'', CERN/LHCC {\bf 1997-032} (1997).

\bibitem{tracker}
CMS Coll., ``Tracker Technical Design Report'', CERN/LHCC {\bf 1998-006} (1998);
Addendum CERN/LHCC {\bf 2000-016} (2000).

\bibitem{ecal}
CMS Coll., ``ECAL Technical Design Report'', CERN/LHCC {\bf 1997-033} (1997).

\bibitem{hcal}
CMS Coll., ``HCAL Technical Design Report'', CERN/LHCC {\bf 1997-031} (1997).

\bibitem{trigger}
CMS Coll., ``L1 Trigger Technical Design Report'', CERN/LHCC {\bf 2000-038} (2000).

\bibitem{daqtdr} CMS Coll., 
``Data Acquisition and High-Level Trigger Technical Design Report'', 
CERN/LHCC {\bf 2002-026} (2002).

\bibitem{lhc} http://lhc-commissioning.web.cern.ch/lhc-commissioning/

\bibitem{ptdr1} CMS Coll., 
``Physics Technical Design Report Vol. 1: Detector Performance and Software'', 
CERN/LHCC {\bf 2006-001} (2006).

\bibitem{ptdr2} CMS Coll., 
``Physics Technical Design Report Vol. 2: Physics Performance'', 
CERN/LHCC {\bf 2006-021} (2006).


\bibitem{rtag} D.~Acosta \etal, 
\cmsnote{2006}{095}. 

\bibitem{misaliscen}
I.~Belotelov \etal, 
\cmsnote{2006}{008}.

\bibitem{misalinote}
P.~Vanlaer \etal, 
\cmsnote{2006}{029}.

\bibitem{hipnote}
V.~Karimaki \etal, 
\cmsnote{2006}{018}.

\bibitem{millenote}
P.~Schleper \etal, 
\cmsnote{2006}{011}.

\bibitem{kalmannote}
R.~Fruehwirth \etal, 
\cmsnote{2006}{022}.

\bibitem{muonalinote}
A.~Calderon \etal, 
\cmsnote{2006}{016}.

\bibitem{ecalcalib} L.~Agostino \etal, 
\cmsnote{2006}{021}.

\bibitem{top1}
M.~Davids \etal, 
\cmsnote{2006}{077}.

\bibitem{top2}
J.~D'~Hondt \etal, 
\cmsnote{2006}{066}.

\bibitem{btagnote}
C.~Weiser, 
\cmsnote{2006}{014}.

\bibitem{bsmumunote}
C.~Eggel \etal, 
\cmscr{2006}{071}.



\end{thebibliography}
\end{document}